\documentclass[lettersize,journal]{IEEEtran}
\usepackage{amsmath,amsfonts}
\usepackage{algorithmic}
\usepackage{algorithm}
\usepackage{array}
\usepackage{multirow}
\usepackage{xcolor}

\usepackage{textcomp}
\usepackage{stfloats}
\usepackage{url}
\usepackage{verbatim}
\usepackage{graphicx}
\usepackage{cite}
\usepackage{caption} 
\usepackage{amsmath}
\usepackage{xspace}
\usepackage{fixltx2e}
\usepackage{subcaption}

\graphicspath{{Figure/}}

\begin{document}

\title{High Performance Signal Design for Optical OFDM Systems using Variational Autoencoder}

\author{Nam N. Luong, Chuyen T. Nguyen, Thanh V. Pham,~\IEEEmembership{Senior Member,~IEEE}

\thanks{Nam N. Luong and Chuyen T. Nguyen are with the School of Electrical and Electronic Engineering, Hanoi University of Science and Technology, Hanoi, Vietnam (email: luongnamnd220503@gmail.com, chuyen.nguyenthanh@hust.edu.vn).

Thanh V. Pham with the Department of Mathematical and Systems Engineering, Shizuoka University, Shizuoka, Japan (e-mail: pham.van.thanh@shizuoka.ac.jp). 
}}

\maketitle

\begin{abstract}
This letter proposes a design of low peak-to-average power ratio (PAPR), low symbol error rate (SER), and high data rate signal for optical orthogonal frequency division multiplexing (OFDM) systems. The proposed design leverages a variational autoencoder (VAE) incorporating gradual loss learning to jointly optimize the geometry and probability of the constellation's symbols. This not only enhances mutual information (MI) but also effectively reduces the PAPR while maintaining a low SER for reliable transmission. We evaluate the performance of the proposed VAE-based design by comparing the MI, SER, and PAPR against existing techniques. Simulation results demonstrate that the proposed method achieves a considerably lower PAPR while maintaining superior  SER and MI performance for a wide range of SNRs.
\end{abstract}


\section{Introduction}
\IEEEPARstart{O}{ptical} wireless communications (OWC) has emerged as one of the most rapidly developing communication technologies in recent years. Due to distinct advantages such as unlicensed spectrum and immunity to electromagnetic interference, OWC is a promising complement and/or alternative to radio frequency (RF) communications \cite{Celik2023}. OWC systems are typically categorized into free-space optics (FSO) and visible light communications (VLC), where the former is usually used for outdoor applications, while the latter is often employed for indoor scenarios. 

To address challenges commonly encountered in OWC systems, such as frequency selective fading and multipath fading, orthogonal frequency division multiplexing (OFDM) has been widely adopted due to its robustness to these impairments. Conventional OFDM cannot be directly applied to OWC systems due to the real and nonnegative signal required by 
%because its symbols are complex-valued and bipolar, whereas 
optical intensity modulation and direct detection (IM/DD).
%systems require real-valued and non-negative signals. 
Therefore, several OFDM variants have been developed specifically for optical systems, notably direct current biased optical OFDM (DCO-OFDM) and asymmetrically clipped optical OFDM (ACO-OFDM)\cite{SurveyACODCO}. Optical OFDM utilizes Hermitian symmetry to ensure real-valued signals before the inverse fast Fourier transform (IFFT). Different methods are then applied to generate non-negative time-domain signals (e.g., DC biasing in DCO-OFDM). 

A major drawback of OFDM-based systems is the inherent high peak-to-average power ratio (PAPR), which arises due to the summation of multiple subcarriers with different phases, occasionally producing large peaks in the signal waveform. Due to the nonlinearity of power amplifiers at the high-power region, this high PAPR can lead to severe nonlinear signal distortions, which results in reduced system efficiency and degraded symbol error rate (SER) performance. Several techniques have been proposed for PAPR reduction, including selective mapping (SLM) \cite{muller1997ofdm}, partial transmit sequence (PTS) \cite{PTS}, and clipping \cite{ochiai2002performance}. Clipping limits the signal amplitude by truncating it beyond a certain threshold. Although simple, this approach introduces signal distortions, hence deteriorating the SER. In contrast, SLM and PTS mitigate signal peaks by modifying phase factors. While SLM selects the lowest PAPR version among multiple phase-shifted signals, PTS optimizes sub-block phases. Though effective, these methods incur high computational complexity due to exhaustive searches for optimal phase factors.

In recent years, deep learning (DL) has gained significant attention in tackling challenges in the physical layer design, particularly using autoencoders (AEs) to model end-to-end communication systems\cite{DLforphysical,cammerer2020trainable}. In this framework, the encoder represents the transmitter, while the decoder functions as the receiver. Several studies have investigated the potential of using AEs for PAPR reduction in OFDM systems \cite{propose,8971873, HAO2019110,huleihel2020low}. %The findings consistently show that autoencoder-based approaches outperform traditional methods, highlighting their effectiveness and promise to improve communication system performance. 
In \cite{propose}, Kim $et~al.$ proposed a PAPR reduction method using an AE called PRnet. Simulation results demonstrated that this approach achieves superior PAPR reduction compared to conventional methods while maintaining a good bit error rate (BER) performance. Using a similar approach, \cite{8971873} employs an AE combined with clipping to reduce PAPR in DCO-OFDM systems. Due to the AE's powerful signal reconstruction capability, the effect of clipping in DCO-OFDM on the BER was minimized. The authors in \cite{HAO2019110} proposed a hybrid AE, combining an AE with a phase rotator to reduce PAPR for ACO-OFDM. This design achieved better results than the AE-only approach at the expense of significantly higher computational complexity. In \cite{huleihel2020low}, the authors designed an OFDM signal with a low PAPR waveform while simultaneously addressing the amplifier nonlinearity.

%Compared with other optical OFDM variants, ACO-OFDM suffers from both reduced achievable data rate (since only a half number of subcarriers is used for carrying data) and higher PAPR \cite{Bai2021}. Hence, methods for improving the performance of ACO-OFDM are of interest. 
In this study, we propose a novel approach for generating OFDM signals with low PAPR, lower SER, and high data rate for optical OFDM systems by leveraging a variational autoencoder (VAE) to optimize the probability and geometry of the constellation's symbols, i.e., probabilistic and geometric constellation shaping (PCS and GCS). Unlike existing methods that consider uniform signaling, the proposed design, by properly shaping the symbol constellation, not only reduces the PAPR but also achieves lower SER and higher data rate which is calculated as the symbol-wise mutual information (MI) \cite{cammerer2020trainable}. To achieve better learning, a two-stage training of the proposed model is performed. In the first stage, the loss function only considers MI to improve the accuracy of symbol detection. Once the first stage is completed, the second stage introduces an additional component of the weighted PAPR. This technique, known as gradual loss learning \cite{huleihel2020low}, helps prevent the network from being \textit{distracted} by multiple competing objectives during training. 
%offers more benefits than conventional methods that only consider PAPR reduction with uniform signaling. 
%By calculating mutual information on a symbol-wise basis \cite{cammerer2020trainable}, we evaluate SER instead of BER as in these previous works. 
%Furthermore, we analyze the impact of our method by examining constellation diagrams across different SNR levels and benchmarking PAPR reduction against traditional techniques.
The main contributions in the paper are summarized as follows.

\begin{itemize}
    \item  We propose a VAE-based signal design for optical OFDM systems that achieves low PAPR, low SER, and high data rate. %ACO-OFDM signals by utilizing a VAE model whose architecture is redesigned to better achieve multi-objective optimization. 
    Specifically, the proposed design employs three neural networks (NNs) for three different tasks: one for PCS optimization, one for GCS optimization and PAPR reduction (i.e., implementing the modulator), and one for signal reconstruction at the receiver (i.e., implementing the demodulator), respectively.
    \item We present comprehensive comparisons regarding PAPR, SER, and MI between the proposed approach and existing methods such as SLM, clipping, and the PRnet \cite{propose}.
\end{itemize}
\textit{Notation}: $H(\cdot)$, $I(\cdot, \cdot)$, and $\mathbb{E}[\cdot]$ are used to denote the entropy, mutual information, and expected operation, respectively. 

\section{System model and Autoencoder}
\subsection{System model}
\begin{figure}[!t]
    \centering
    \includegraphics[scale = 0.3]{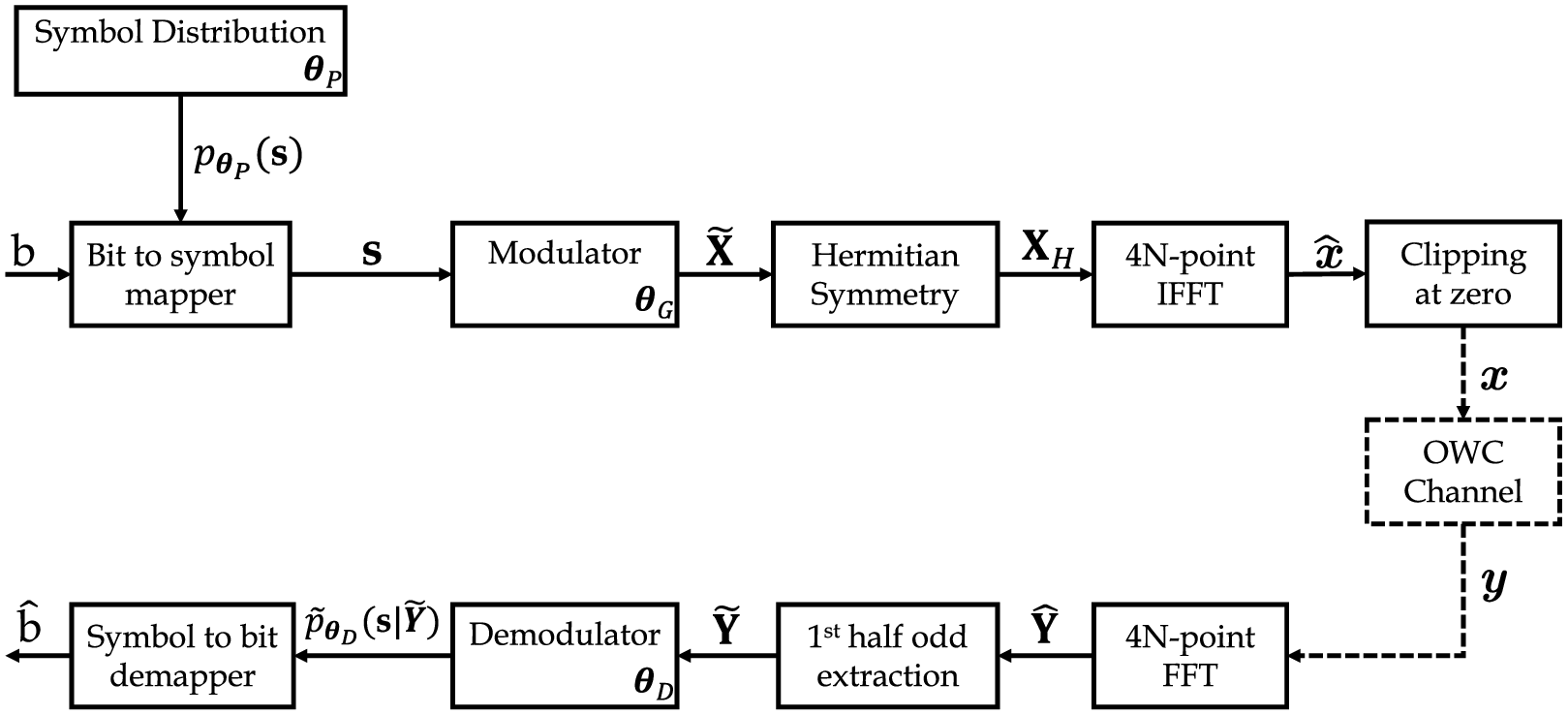}
    \caption{Schematic diagram of ACO-OFDM systems.}
    \label{fig:aco_ofdm}
    %\vspace{-0.5cm}
\end{figure}
%Paragraph 1: Give the abstract of ACO OFDM.
For the sake of conciseness, in the following, we present the proposed method for the case of ACO-OFDM\footnote{The same analysis can be readily applied to DCO-OFDM, which is not included due to the page limitation.}.  As illustrated in Fig.~\ref{fig:aco_ofdm}, we consider in this study a single-input single-output (SISO) ACO-OFDM system employing intensity modulation/direct detection (IM/DD) and $M$-ary quadrature amplitude modulation ($M$-QAM) with $4N$ subcarriers where the input bitstream \(b\) is mapped to a set QAM symbols $\mathbf{s}$ by a bit-to-symbol mapping block according to a parametric distribution $p_{\mathbf{\theta}_P}(\mathbf{s})$ with parameters $\pmb{\theta}_P$ which is trained by a NN. 
%This block consists of essential operations such as serial-to-parallel (S/P) conversion and $M$-QAM mapping, producing the output symbols \(\tilde{X}\). 
For simplicity and clarity, in the subsequent sections, we assume the presence of this mapping block and focus on the symbols rather than the input bitstream. 
%Subsequently, the symbols are rearranged using the Hermitian symmetry operation, where only one-quarter of the total subcarriers (\(N\) odd subcarriers) are utilized to carry the symbols \(\tilde{X}\). Another quarter of the subcarriers is assigned to the conjugates of \(\tilde{X}\), while the remaining half ($2N$ even subcarriers) is allocated to null symbols. 
The symbols are then fed into a modulator, which is implemented as a NN with trainable parameters $\pmb{\theta}_G$, which reconstructs \textcolor{black}{the transmitted symbols.} 
To ensure a real time-domain signal after IFFT, the symbols are then rearranged using the Hermitian symmetry operation, where one-quarter of the total subcarriers, specifically the odd-indexed subcarriers, carry the data signals $\tilde{\mathbf{X}}=[\tilde{X}_0,\tilde{X}_1,\ldots,\tilde{X}_{N-1}]$. Another quarter is assigned to the conjugates of \(\tilde{X}\), while the remaining half, corresponding to the even-indexed subcarriers, is allocated to null data signals. Accordingly, the frequency-domain data vector is given by $\mathbf{X}_H=[0, \tilde{X}_0,0,\tilde{X}_1,\ldots,0,\tilde{X}_{N-1},0,\tilde{X}_{N-1}^*,\ldots,0,\tilde{X}^*_1,0,\tilde{X}^*_0]$ 
where \(\tilde{X}^*_n\) denotes the complex conjugate of \(\tilde{X}_n\). %After applying Hermitian symmetry, the frequency-domain signal \(X_H\) is transformed into the time domain through a \(4N\)-point Inverse Fast Fourier Transform (IFFT). This step ensures that the resulting 
The real time-domain signal \(\hat{x}(n)\) is then generated through an 4N-point IFFT as follows
\begin{align}
&\hat{x}(n) = \frac{1}{\sqrt{4N}} \sum_{k=0}^{4N-1} \mathbf{X}_H(k) e^{\frac{j2\pi nk}{4N}},~n = 0, 1, \dots, 4N-1.
\end{align}
Since the IM/DD scheme requires the transmit signal to be nonnegative, clipping at 0 is applied to $\hat{x}(n)$ to remove all negative signals.
%To ensure the signal remains suitable for transmission by intensity modulation/direct detection methods (IM/DD), amplitude clipping is applied to remove negative signal values, as the light-emitting diode (LED) can only transmit nonnegative signals. 
The clipped signal, \(x(n)\), is  defined as

\begin{align}
x(n) = 
\begin{cases} 
0, & \hat{x}(n) < 0 \\
\hat{x}(n), & \hat{x}(n) \geq 0
\end{cases}\ 
\end{align}
Due to the particular symbol arrangement in frequency-domain data vector $\mathbf{X}_H$, the time-domain signal \(\hat{x}\) becomes symmetric around zero. Therefore, clipping the negative-valued portions of the signal does not lead to any information loss. 
As a result, unlike DCO-OFDM, ACO-OFDM does not require an additional DC bias to enable a nonnegative signal, thus being more power efficient. 

OWC channels can be reasonably modeled as quasi-static channels (in case of systems for indoor scenarios, e.g., VLC systems) or block-fading channels (in case of systems for outdoor scenarios, e.g., FSO systems). Since the channel coherence time is much longer than the ODFM symbol duration, without loss of generality, the channel gain is normalized to unity during the transmission of one symbol. Thus, the received time-domain signal can be written as 
\begin{align}
    y(n) = x(n) + z(n),
\end{align}  
where $z(n)$ is assumed to be Gaussian noise with zero-mean and variance $\sigma^2$. It should be noted that in certain scenarios, the discrete-time Poisson noise model (when the received optical power is too low) or the signal-dependent noise model (when the received optical power is substantially higher than that of background light) characterizes the channel more accurately \cite{Chaaban2020}. Though it is interesting to investigate these channel models, we restrict our attention to the optimal constellation problem under the simple Gaussian noise model. After the FFT, denote $\tilde{\mathbf{Y}}$ as the data-carrying frequency-domain signal vector. 
The demodulator is also implemented as an NN with trainable parameters $\pmb{\theta}_D$, which reconstructs the symbol $\mathbf{s}$ from  $\tilde{\mathbf{Y}}$ using the mapping $\tilde{p}_{\pmb{\theta}_D}(\mathbf{s}|\tilde{\mathbf{Y}})$. The sent bits are then recovered from the symbol-to-bit demapper. 

One of the main drawbacks of ACO-OFDM, as well as OFDM in general, is its high PAPR, which is expressed by
\begin{align}
\text{PAPR}\{x(n)\} = \frac{\underset{0 \leq n\leq 4N-1}{\max}|{x}(n)|^2}{\mathbb{E}[|{x}(n)|^2]}.
\label{papr-formula}
\end{align}
Here, the dominator denotes the average power of ${x}(n)$.
%In order to reduce PAPR, there are various methods employed, such as selective mapping (SLM) and clipping. 
%The comparison between these methods is typically visualized using 
The severity of PAPR is often characterized by the complementary cumulative distribution function (CCDF), which implies the probability that the signal PAPR exceeds a predefined threshold $\text{PAPR}_0$, i.e., $\text{Pr($\text{PAPR}$ $\geq$ $\text{PAPR}_0$)}$.
\subsection{Modeling End-to-End Systems as AEs}
A general AE architecture consists of two main blocks: an encoder $f(x)$ and a decoder $h(x)$, where $x$ denotes the input data. It is trained to minimize a certain loss function denoted as $\mathcal{L}(x, h(f(x))$. Using AEs to optimize the design of communication systems has received considerable interest over the past few years since an end-to-end communication system can be naturally interpreted as an AE in which the transmitter and receiver can be viewed as the encoder and decoder of the AE \cite{stark2019joint,aoudia2020joint}. 

For the considered ACO-OFDM system depicted in Fig.~\ref{fig:aco_ofdm}, the encoder at the transmitting side performs PCS and GCS through two NNs with trainable parameters $\pmb{\theta}_P$ and $\pmb{\theta}_G$, respectively. Let $p_{\pmb{\theta}_P}(\mathbf{s})$ and $g_{\pmb{\theta}_G}(\mathbf{s})$ be the probabilistic shaping distribution and the geometric shaping function. The distribution of the ACO Tx input $\tilde{\mathbf{X}}$ is then given by  
%Recently, autoencoders have emerged as a powerful tool to enhance performance at the physical layer. Specifically, the use of autoencoders for shaping problems has been proven to be effective, as most of the results on mutual information achieved by autoencoders outperform traditional methods 
%Autoencoders are employed to shape both the geometry and probability of the constellation's symbols, based on the following principles
\begin{align}
p_{\pmb{\theta}_P, \pmb{\theta}_G}(\tilde{\mathbf{X}})=\sum_{n=0}^{N-1}\sum_{i=1}^{M}\delta(\tilde{X}_n-g_{\pmb{\theta}_G}(\mathbf{s}_i))p_{\pmb{\theta}_P}(\mathbf{s}_i),
\end{align}
where $\delta(\cdot)$ denotes the Dirac function.
%$p_{\theta_P}$ and $g_{\theta_G}$ perform geometric shaping and probabilistic shaping, which can be done by the neural networks with trainable parameters $\theta_P$ and $\theta_G$, respectively. 
% The goal of constellation shaping is to find $p_{\theta_P,\theta_G}(x)$ that maximizes mutual information between channel input and channel output.
% \begin{align}
% I(X,Y) = \mathbb{E}\left[\log_2\frac{p(y|x)}{\sum_{x \in X}p(y|x)p_{\theta_P,\theta_G}(x)}\right]
% \end{align}
% where $x, y$ are the realizations of the channel input $X$ and channel output $Y$, respectively.
At the receiving side, the demodulator, which is also modeled as an NN with trainable parameter $\pmb{\theta}_D$, performs a classification task (i.e., detection) based on the ACO Rx output $\tilde{\mathbf{Y}}$. As a result, the categorical cross entropy can be used as a suitable loss function \cite{stark2019joint}
\begin{align}
\mathcal{L}(\pmb{\theta}_P,\pmb{\theta}_G,\pmb{\theta}_D)&=\mathbb{E}\left[-\log(\tilde{p}_{\pmb{\theta}_D}(\mathbf{s}|\tilde{\mathbf{Y}}))\right] \nonumber \\&= H_{\pmb{\theta}_P}(\mathbf{s})-I_{\pmb{\theta}_P,\pmb{\theta}_G}(\tilde{\mathbf{X}},\tilde{\mathbf{Y}})\nonumber\\& \quad +  \mathbb{E}\left[D_{\text{KL}}(p_{\pmb{\theta}_P,\pmb{\theta}_G}(\tilde{\mathbf{X}}|\tilde{\mathbf{Y}})~||~\tilde{p}_{\pmb{\theta}_D}(\tilde{\mathbf{X}}|\tilde{\mathbf{Y}}))\right],
\label{loss-function1}
\end{align}
where $\tilde{p}_{\pmb{\theta}_D}(\mathbf{s}|\tilde{\mathbf{Y}})$ represents an approximation of the true mapping $p_{\pmb{\theta}_P,\pmb{\theta}_G}(\mathbf{s}|\tilde{\mathbf{Y}})$, {\color{black}which maps the received output $\tilde{\mathbf{Y}}$ to a probability vector over the set of symbols denoted as $\mathbf{s}$} and $D_{\text{KL}}(\cdot~||~\cdot)$ is the Kullback-Leibler divergence, which measures the difference between two distributions. However, two issues exist with the above-defined loss function. Firstly, minimizing \eqref{loss-function1}  corresponds to minimization of the source entropy $H_{\pmb{\theta}_P}(\mathbf{s})$, which in turn reduces the achievable data rate. Secondly, the PAPR is not taken into consideration. Note that, in our proposed design, the PAPR given in \eqref{papr-formula} is influenced by the GCS, thus, is parameterized by $\pmb{\theta}_G$. 
%, $\theta_D$ is the parameters of the demodulator. This type of loss function is characteristic of a variational autoencoder (VAE). 
To circumvent these issues, we consider the following modified loss function for training 
\begin{align}
\tilde{\mathcal{L}}(\pmb{\theta}_P,\pmb{\theta}_G,\pmb{\theta}_D)=\mathcal{L}(\pmb{\theta}_P,\pmb{\theta}_G,\pmb{\theta}_D)-H_{\pmb{\theta}_P}(\mathbf{s}) + \lambda\text{PAPR}_{\pmb{\theta}_G},
\label{eq:loss-function}
\end{align}
where $\lambda$ is a positive factor that determines the dominance of PAPR on the overall loss. When the factor value is high, the training prioritizes reducing the PAPR and vice versa.

% with the loss function designed as above, minimizing $\tilde{\mathcal{L}}$ is equivalent to maximize $I(X,Y)$, while minimizing the KL divergence between the approximation $\tilde{p}_{\pmb{\theta}_D}(x|y)$ and the true posterior $p_{\pmb{\theta}_P,\pmb{\theta}_G}(x|y)$. Moreover, symbol error rate (SER) also decreases when maximizing $I(X,Y)$ between transmitter and receiver, we consider SER due to the symbol-wise approach \cite{cammerer2020trainable}.

%Due to the task of reducing PAPR, we add a additional component of weighted PAPR in the loss function for training. PAPR is reduced by constellation modification which is directly implemented by NN2 with parameters $\pmb{\theta}_G$.
% \begin{align}
% \hat{\mathcal{L}}(\pmb{\theta}_P,\pmb{\theta}_G,\pmb{\theta}_D) = \mathcal{L}_1
% +\lambda\mathcal{L}_2
% \label{loss_function}
% \end{align}
% here $\mathcal{L}_1=\tilde{\mathcal{L}}$
% defined in (\ref{eq:tildeL}), $\mathcal{L}_2 = \text{PAPR}_{\pmb{\theta}_G}$, and $\lambda$ is a positive factor that determines which loss $\mathcal{L}_1$ or $\mathcal{L}_2$ is dominant. When the factor value is high, the model puts more effort into reducing PAPR and vice versa.

\section{Training End-to-End System}
% \clearpage
\begin{figure}[ht]
    \centering
    \includegraphics[width = \linewidth, height = 0.35\linewidth]{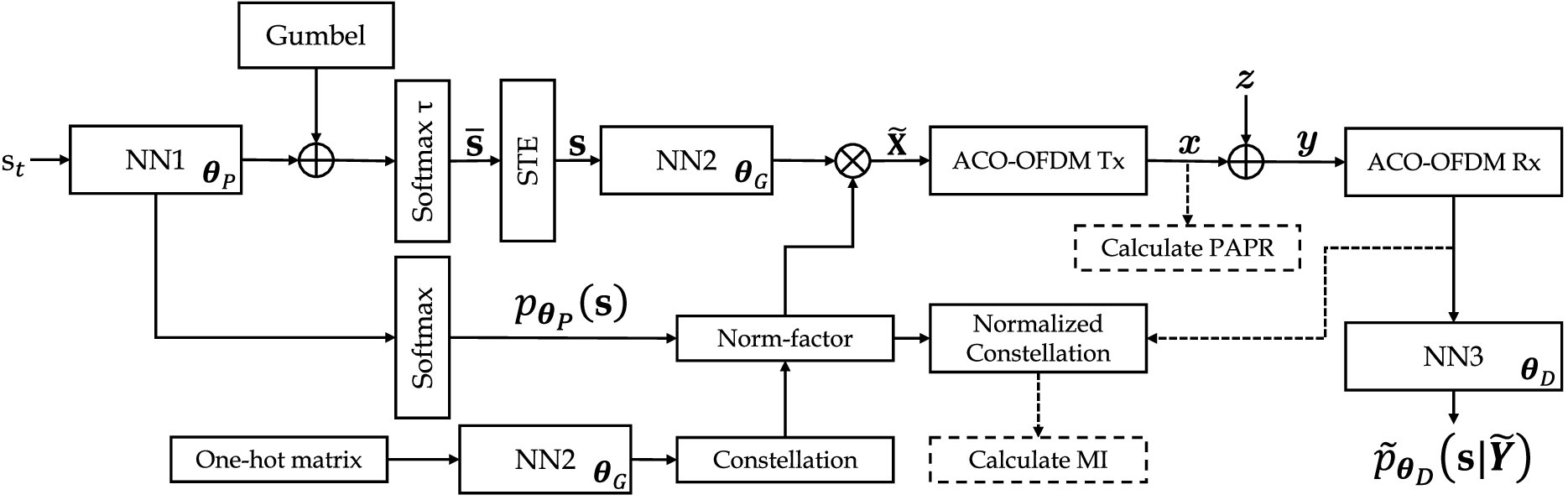}
    \caption{Proposed architecture of end-to-end system.}
    \label{fig:end-to-end system}
    %\vspace{-0.2cm}
\end{figure}
% \clearpage
\subsection{Proposed Structure of End-To-End System}
%This section illustrates how the VAE implements probabilistic and geometric shaping. 
Figure \ref{fig:end-to-end system} illustrates the proposed end-to-end system architecture. As discussed in \cite{stark2019joint}, performing PCS using NNs is challenging due to the discreteness of the symbol distribution $p_{\pmb{\theta}_P}(\mathbf{s})$, which renders training NNs using gradient-based optimization methods difficult. 
%difficulty of training a sampling mechanism for a symbol $s$ drawn from the set $\mathcal{S}$.   %Firstly, the PCS can be done by sampling the distribution of symbols $s$ from the set of $S$. However, since \( s \) follows a discrete distribution, it is not differentiable through the sampling block, posing challenges for gradient-based optimization. 
To overcome this issue, we employ the Gumbel-Softmax trick \cite{jang2017categorical}, which samples a symbol $\mathbf{s}_i$ from the distribution $p_{\pmb{\theta}_P}(\mathbf{s})$ as follows
%which is highly effective in addressing the challenges of categorical reparameterization. This approach not only allows for differentiable sampling but also seamlessly integrates into gradient-based learning frameworks, making it a powerful tool for training end-to-end systems that require discrete symbol generation.
\begin{align}
\mathbf{s}_i=\underset{i=1,\ldots,M}{\operatorname{arg~max}}(g_i+\log(p_{\pmb{\theta}_P}(i))),
\end{align}
where $g_i$ are independent and identically distributed samples drawn from the standard Gumbel(0,1) distribution\footnote{If  $u\sim\mathcal{U}(0,1)$ be a uniform random variable over (0, 1), then $-\log(-\log(u)) \sim \text{Gumbel}(0, 1)$.}. The softmax function is then used as a continuous, differentiable approximation to the $\operatorname{arg~max}$ operator to generate $M$-dimensional sample vectors $\bar{\mathbf{s}}$ where each element of the vector is given by
\begin{align}
\bar{\mathbf{s}}_i=\operatorname{softmax}((\log(p_{\pmb{\theta}_P}(i))+g_i)/\tau)\ \text{for}\ i=1,\ldots, M.
\end{align}
Here, the positive parameter $\tau$ is called temperature. As the temperature $\tau$ approaches 0, the samples generated by the Gumbel-Softmax method approximate a one-hot vector, and their distribution becomes identical to the categorical distribution $p_{\pmb{\theta}_P}(\mathbf{s})$\cite{jang2017categorical}. In our proposed end-to-end system, the first neural network (i.e., NN1), which takes the pre-known signal-to-noise ratio (SNR) as input, is utilized to generate the \textit{logits} of the symbol distribution $p_{\pmb{\theta}_P}(\mathbf{s})$. The distribution $p_{\pmb{\theta}_P}(\mathbf{s})$ can then be obtained by applying a softmax function to the \textit{logits}.

Compared to \cite{stark2019joint}, our proposed model architecture has a key difference due to the PAPR reduction problem. Instead of multiplying the constellation with the $M \times M$ one-hot matrix obtained from the sampling step to choose the constellation point for symbols, we use a neural network (i.e., NN2) to directly map the $M \times M$ one-hot matrix to the $M \times 2$ matrix containing the real and image of the constellation points and do GCS at the same time. Thus, the NN2 learns how to map each one-hot vector to a constellation point. Afterward, this NN is reused to directly map $\mathbf{s}$ to a set of unnormalized constellation points before going to the ACO-OFDM transmitter. With this design, the PAPR can be effectively controlled by NN2. Without loss of generality, assume that the average symbol power is normalized to unity, i.e., $\mathbb{E}\left[|\tilde{X}_n|^2\right] = 1$. To guarantee this normalization when performing PCS and GCS, 
the data signal $\tilde{\mathbf{X}}$ is obtained by multiplying the unnormalized constellation points by a normalization factor $\gamma$, which is calculated as follows 
% The modulator, which performs geometric shaping, is implemented as an NN different from the first one in the sampling stage. This NN takes a \(M\times M\) one-hot matrix as input, and output is a \(M\times 2\) matrix which contains the real parts and image parts of the constellation points. This implies that the NN is a mapping operation that transforms one-hot vectors into constellation points directly. Subsequently, it can be seen in Fig.\ref{fig:set up for training}, this NN is utilized to convert the one-hot vectors \( s \) into constellation points. As a result, it can influence the PAPR more effectively compared to merely multiplying one-hot matrices to select constellation points, as done in\cite{stark2019joint}. Normalization step is to meet the requirement of expected energy equals one, i.e.,
% \begin{align}
% \sum_{s \in S}p_{\theta_P}(s)|x_G|_2^2=1.
% \label{10}
% \end{align}
\begin{align}
\gamma = \frac{1}{|x_G|\sqrt{\sum_{s\in \mathbf{s}}p_{\pmb{\theta}_P}(\mathbf{s})}},
\end{align}
where $x_G$ is the geometric constellation points (i.e., the output of the NN2). 
%as well as the output of NN2 when mapping the $M \times M$ one hot matrix to the $M \times 2$ matrix, i.e., GCS. Before being fed into the ACO-Tx, the data signal $\tilde{\mathbf{X}}$ is multiplied by $\gamma$ to meet the unit energy requirement. In addition, the normalized constellation also serves as a reference for MI calculation, and all of the process is presented in Fig.\ref{fig:end-to-end system}.
Note that, in the sampling process, \(\bar{\mathbf{s}}\) is merely an approximation of the vector \(\mathbf{s}\). Meanwhile, the input to NN2 is the true one-hot vector matrix. Therefore, a straight-through estimator (STE) is required \cite{bengio2013estimating}, which uses the true one-hot vector \(\mathbf{s}\) for the forward pass and the approximation of \(\mathbf{s}\) for the backward pass (backpropagation) at training.

The demodulator is an NN (i.e., NN3), which is made up of fully-connected layers followed by ReLU activation functions. Its output is passed through a softmax function to obtain the probability distribution of symbol vector \( \mathbf{s} \). All neural networks in the system take SNR as input to adapt their parameters, which is defined as  
\begin{align}
\text{SNR}=\frac{\mathbb{E}[x^2(n)]}{\mathbb{E}[z^2(n)]}=\frac{\mathbb{E}[x^2(n)]}{\sigma^2}.
\end{align}
Here, $\mathbb{E}[x^2(n)] = \frac{1}{2}\frac{2\sum_{n=0}^{N-1}\mathbb{E}\left[|\tilde{X}_n|^2\right]}{4N}$ is the average electrical power of $x(n)$. Since the average symbol power is normalized to unity, one has $\mathbb{E}[x^2(n)] = \frac{1}{4}$.
%, which depends on the average electronic energy per QAM symbol  (i.e., $\mathbb{E}[|x(n)|^2]=\frac{\mathbb{E}[|\tilde{X}|^2]}{4}=\frac{1}{4}$).}
%, ensuring maximum {\color{blue}$I(\tilde{\mathbf{X}},\tilde{\mathbf{Y}})$} is achieved at the given SNR.%
\subsection{Training}
% The main objective of constellation shaping is to find \( p_{\theta_P}(s) \) such \( I(X,Y) \) is maximized. Therefore, to perform constellation shaping, (\ref{eq:tildeL}) is considered as the loss function

% Our loss function is designed with two objectives: maximal MI and minimal PAPR. These objectives are performed by two terms $\mathcal{L}_1$ and $\mathcal{L}_2$, respectively 
% \begin{align}
% \mathcal{L}_1=\tilde{\mathcal{L}}
% \end{align}
% \begin{align}
% \hat{\mathcal{L}}(\pmb{\theta}_P,\pmb{\theta}_G,\pmb{\theta}_D) = \mathcal{L}_1
% +\lambda\mathcal{L}_2
% \label{loss_function}
% \end{align}
% To train the model, we use an initial symbol $S_t$ with an arbitrary value as the input of the first NN, here we choose $S_t=1+1j$ and the end-to-end training process needs the pre-known SNR values to tune the optimal constellation. The configuration of each NNs can be seen in table \ref{tab:nn_config}.
% here $\mathcal{L}_1=\tilde{\mathcal{L}}$
% defined in (\ref{eq:tildeL}), $\mathcal{L}_2 = \text{PAPR}$, and $\lambda$ is a positive factor that determines which loss $\mathcal{L}_1$ or $\mathcal{L}_2$ is dominant. When the factor value is high, the model puts more effort into reducing PAPR and vice versa.
The training process for the proposed architecture is divided into two phases. In the first phase, the model is trained using a loss function $\tilde{\mathcal{L}}_1$ without considering the PAPR, i.e., $\tilde{\mathcal{L}}_1(\pmb{\theta}_P,\pmb{\theta}_G,\pmb{\theta}_D)=\mathcal{L}(\pmb{\theta}_P,\pmb{\theta}_G,\pmb{\theta}_D)-H_{\pmb{\theta}_P}(\mathbf{s})$. The purpose of this training phase is to maximize the MI performance $I(\tilde{\mathbf{X}}, \tilde{\mathbf{Y}})$, which leads to more accurate signal reconstruction, hence reducing the SER. At the start of the training, an initial arbitrary symbol $s_t$ is used as the input of the NN1. For our implementation, we choose the symbol $s_t=1+1j$. 
%and the end-to-end training process needs the pre-known SNR values. 
%The configuration of each NNs can be found in Table \ref{tab:nn_config}. 
Once the first training stage is completed, the model is re-trained using the loss function defined in \eqref{eq:loss-function} where $\lambda = 0.01$ is chosen. In the second training phase, the PAPR is directly influenced by NN2, which is responsible for GCS. This neural network adjusts the geometric shape of the constellation so that both SER and PAPR are minimized. 
%The optimal constellation for our multi-objective optimization problem can be found in Fig.\ref{fig:constellation}.
% Fig.\ref{fig:constellation} illustrates the learned constellation with respect to
% the input SNR, the brightness of each constellation point is
% directly proportional to its probability of occurrence. It is
% evident that at low SNR regime, the transmitter focuses on
% transmitting constellation points closer to the center (with
% higher probability) to minimize transmission errors. In con
% trast, in the high SNR regime, the system gradually distributes
% the constellation points more evenly in both the spacing
% and the transmission probability to maximize the transmitted
% information.
\section{Simulation Results and Discussions}
In this section, we present simulation results to illustrate the effectiveness of the proposed design in comparison with existing methods. We consider OWC systems where the symbol duration is typically much shorter than the channel coherent time. As a result, the channel is assumed static during the symbol duration. An optical OFDM system employing square QAM with 64 subcarriers is considered. The NNs of the proposed model are implemented using the PyTorch framework where the training is performed using the Adam optimizer with a batch size of 3000, 30 samples for each training, a learning rate of 0.01, and a fixed temperature of 1. Each NN in the proposed model has one input layer, three hidden layers, and one output layer
The detailed settings of the NNs are presented in TABLE  \ref{tab:nn_config}.
\begin{table}[t]
    \centering
    \caption{Parameters of NNs in the proposed method.}
   \resizebox{0.4\textwidth}{!}{\begin{tabular}{cc|c|c|c|}
        \cline{3-5}
        & & \multicolumn{3}{c|}{\textbf{Number of nodes}} \\ \cline{3-5}
        & & \textbf{NN1} & \textbf{NN2} & \textbf{NN3} \\ \cline{1-5}
        \multicolumn{1}{ |c  }{\multirow{5}{*}{}} & Input layer & 1 & $M$ & 2 \\ \hline
        \multicolumn{1}{ |c  }{} & Hidden layer 1 & 128 & 128 & 128 \\ \hline
        \multicolumn{1}{ |c  }{} & Hidden layer 2 & 512 & 512 & 512 \\ \hline
        \multicolumn{1}{ |c  }{} & Hidden layer 3 & 128 & 128 & 128 \\ \hline
        \multicolumn{1}{ |c  }{} & Output layer & $M$ & 2 & $M$ \\ \hline
        \multicolumn{1}{ |c  }{} & Activation function & ReLU & ReLU & ReLU \\ \cline{1-5}
    \end{tabular}}

    \vspace{-5pt}
    \captionsetup{labelformat=simple, labelsep=colon} 
    
    \label{tab:nn_config}
\end{table}
\begin{figure}[ht]
    \centering
    % First image
    \begin{subfigure}{0.3\linewidth}
        \centering
        \includegraphics[width=\linewidth]{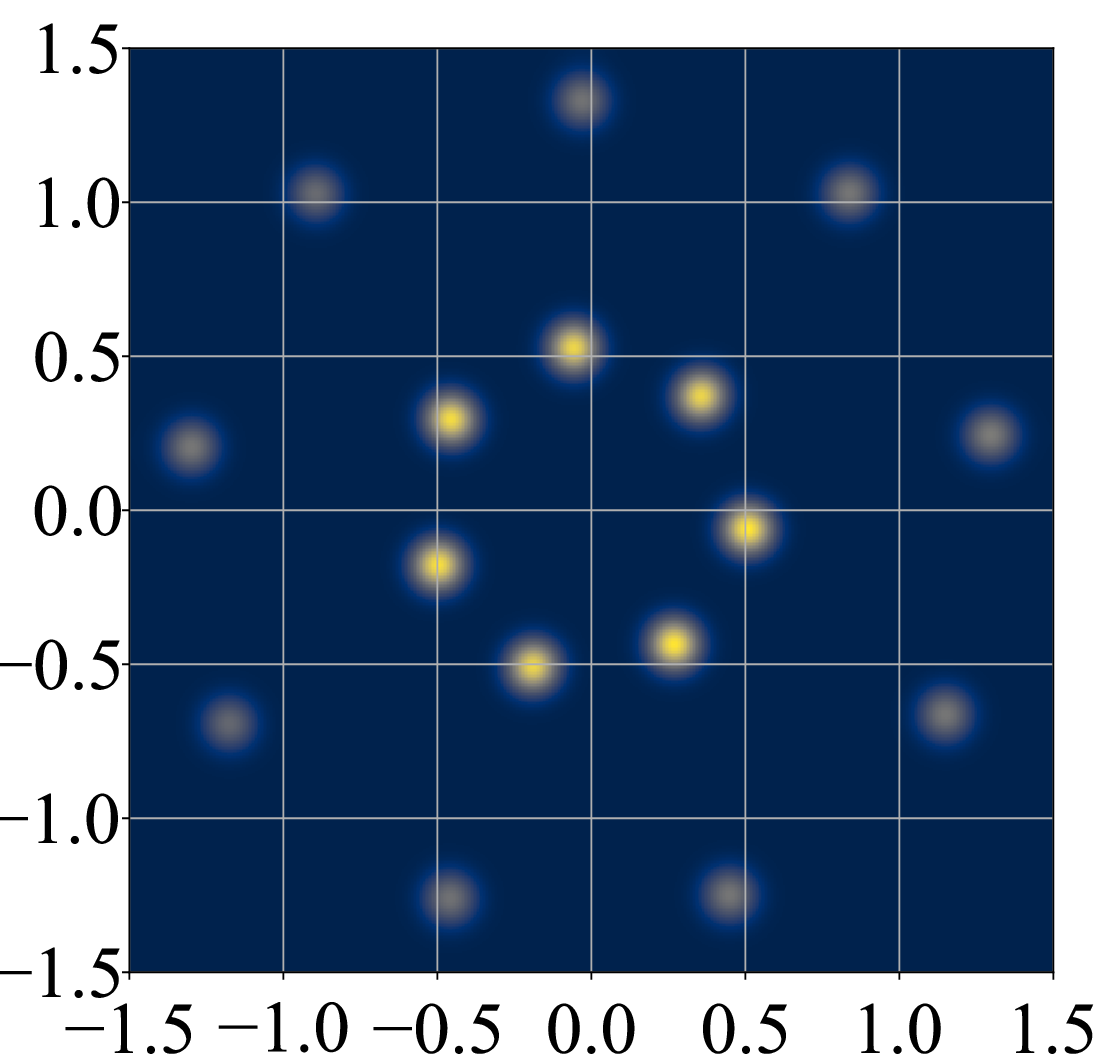}
        \caption{SNR = $5$ dB.}
        \label{fig:5dB_ACO}
    \end{subfigure}
    \hfill
    % Second image
    \begin{subfigure}{0.3\linewidth}
        \centering
        \includegraphics[width=\linewidth]{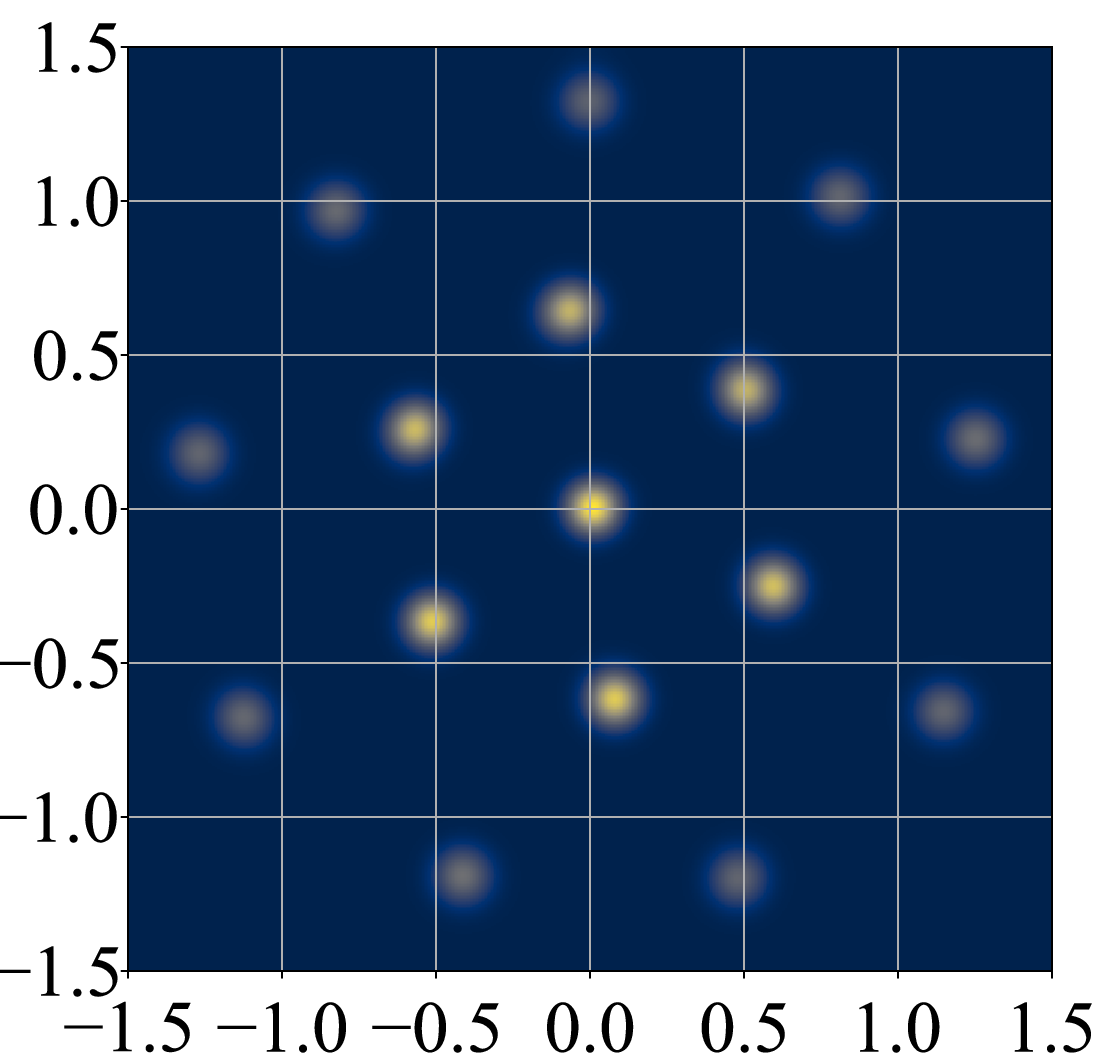}
        \caption{SNR = $10$ dB.}
        \label{fig:10dB_ACO}
    \end{subfigure}
    \hfill
    \begin{subfigure}{0.3\linewidth}
        \centering
        \includegraphics[width=\linewidth]{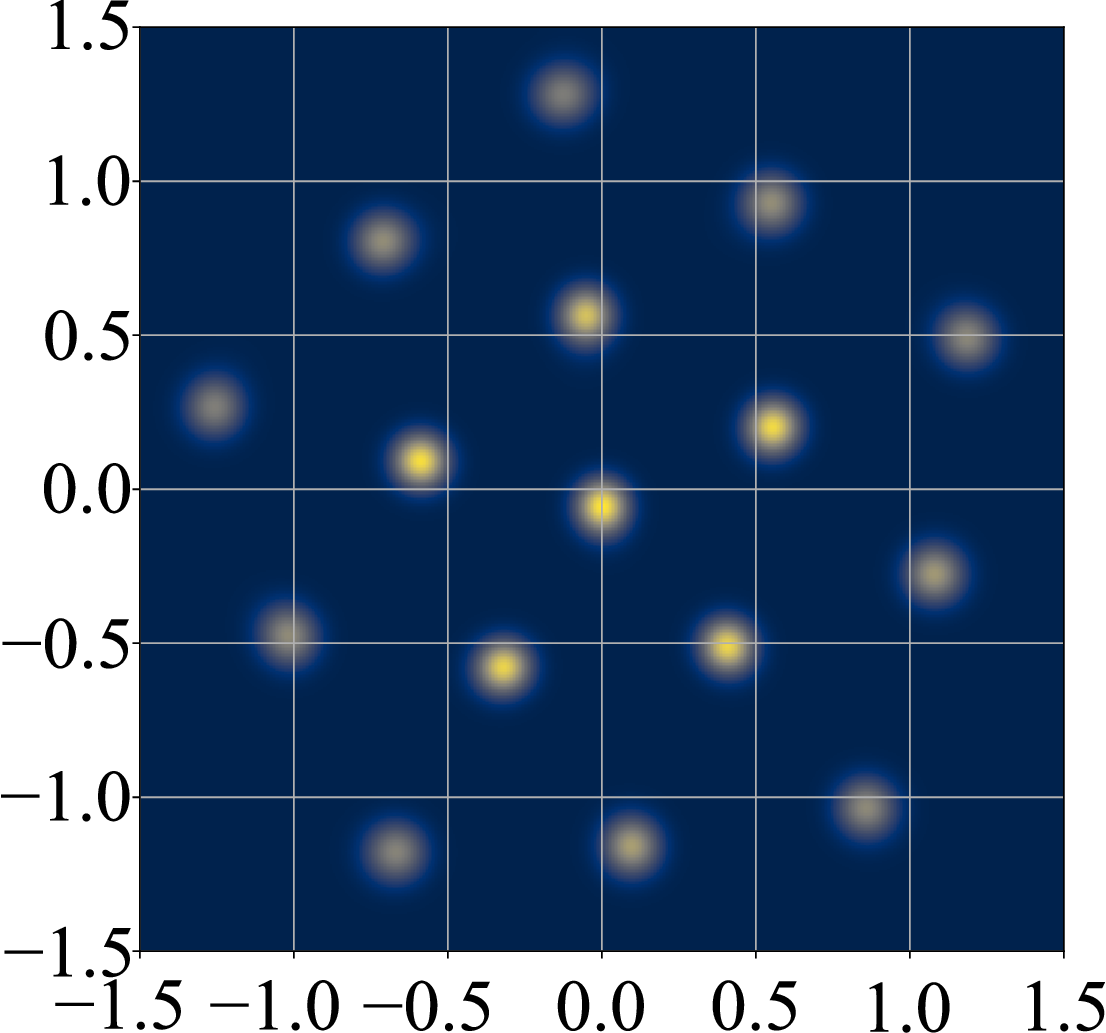}
        \caption{SNR = $15$ dB.}
        \label{fig:15dB_ACO}
    \end{subfigure}
    \caption{Optimal constellations of $16$-QAM for different values of SNR.}
    \label{fig:constellation-ACO}
\end{figure}

\begin{figure}[ht]
    \centering
    \includegraphics[width= .8\linewidth, height = .6\linewidth]{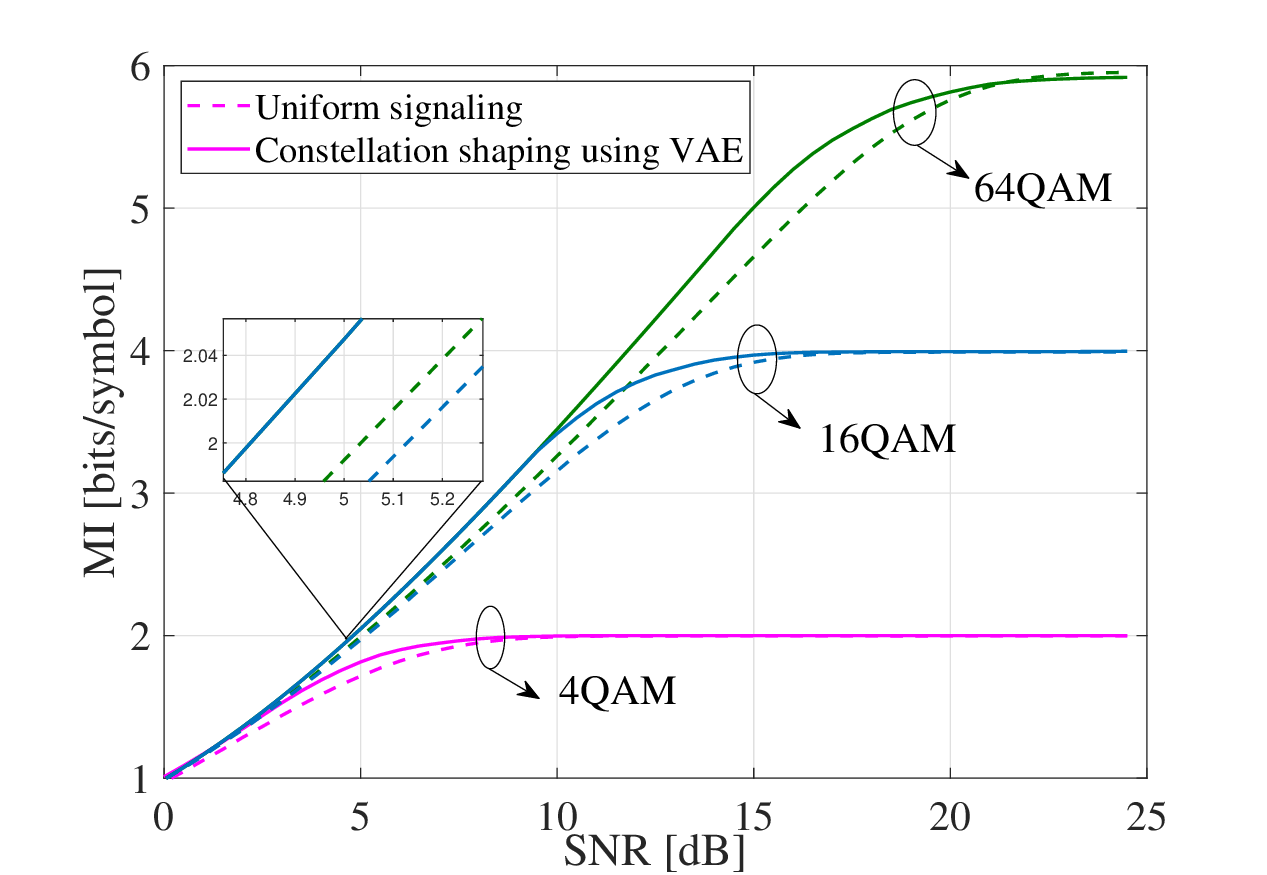}
    \caption{MI of the proposed method in comparison with the uniform signaling.}
    \label{fig:MI-ACO}
\end{figure}

Firstly, Fig.~\ref{fig:constellation-ACO} illustrates the learned 16-QAM constellation with respect to the SNR. The brightness of each constellation point is
proportional to its probability of occurrence. It is shown that the optimal constellations appear different under different SNR regions due to the trade-off between power constraint and robustness to noise. At the low SNR regime (e.g., SNR = 5 dB) where the noise is dominant, lower-energy symbols and high-energy symbols are placed as far as possible to ensure a low SER. Moreover, high-energy symbols are assigned low occurrence probabilities to meet the power constraint. As SNR increases, where symbols are less prone to noise, symbols are spaced more evenly, and their occurrence probabilities tend to be more uniform. This is to maximize the MI.

Fig.~\ref{fig:MI-ACO} compares the MI performance of the proposed VAE-based shaped signaling with the uniform signaling employed by the conventional ACO-OFDM for 4-, 16-, and 64-QAM constellations. 
%illustrates the relationship between MI and SNR (dB) for different modulation schemes, including 4QAM, 16QAM, and 64QAM, with and without shaping. 
%The Shannon capacity limit is also provided as a reference. 
It is clearly shown that our proposed method with proper constellation shaping achieves better MI than the conventional transmission schemes with the uniform constellation. Since the shaped constellation becomes more uniform as the SNR increases (as illustrated in Fig.~\ref{fig:constellation-ACO}), the MI performance of the proposed method approaches that of the uniform signaling. 
%As the SNR increases, the MI performance of the proposed method approaches that of the uniform signaling, which indicates that the optimal constellation becomes more uniform in the high SNR region.  

\begin{figure}[ht]
    \begin{subfigure}[b]{0.48\linewidth}
     \centering
         \includegraphics[width= \linewidth, height = .8\linewidth]{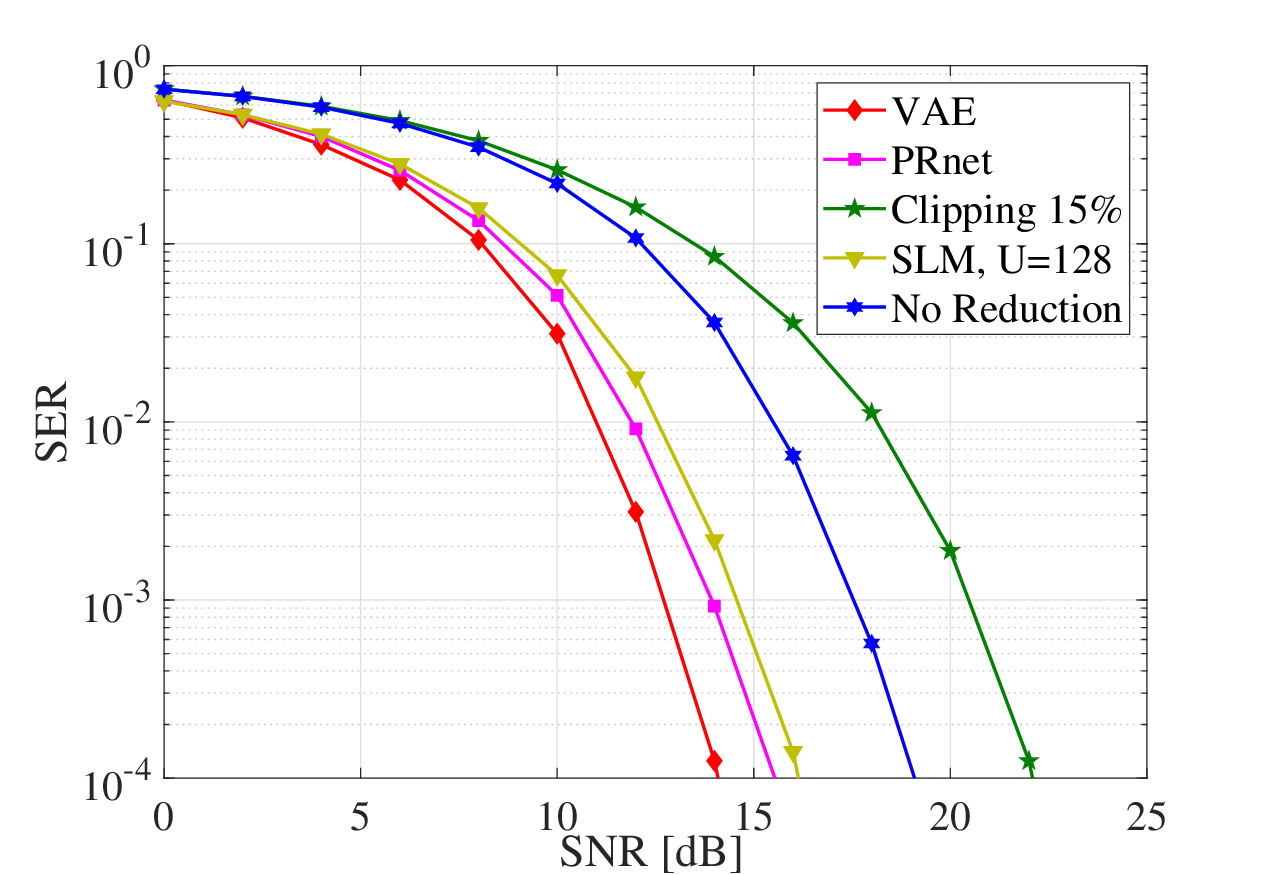}
         \caption{SER.}
         \label{fig:SER-ACO}
     \end{subfigure}
     \begin{subfigure}{0.48\linewidth}
     \centering
     \includegraphics[width= \linewidth, height = .8\linewidth]{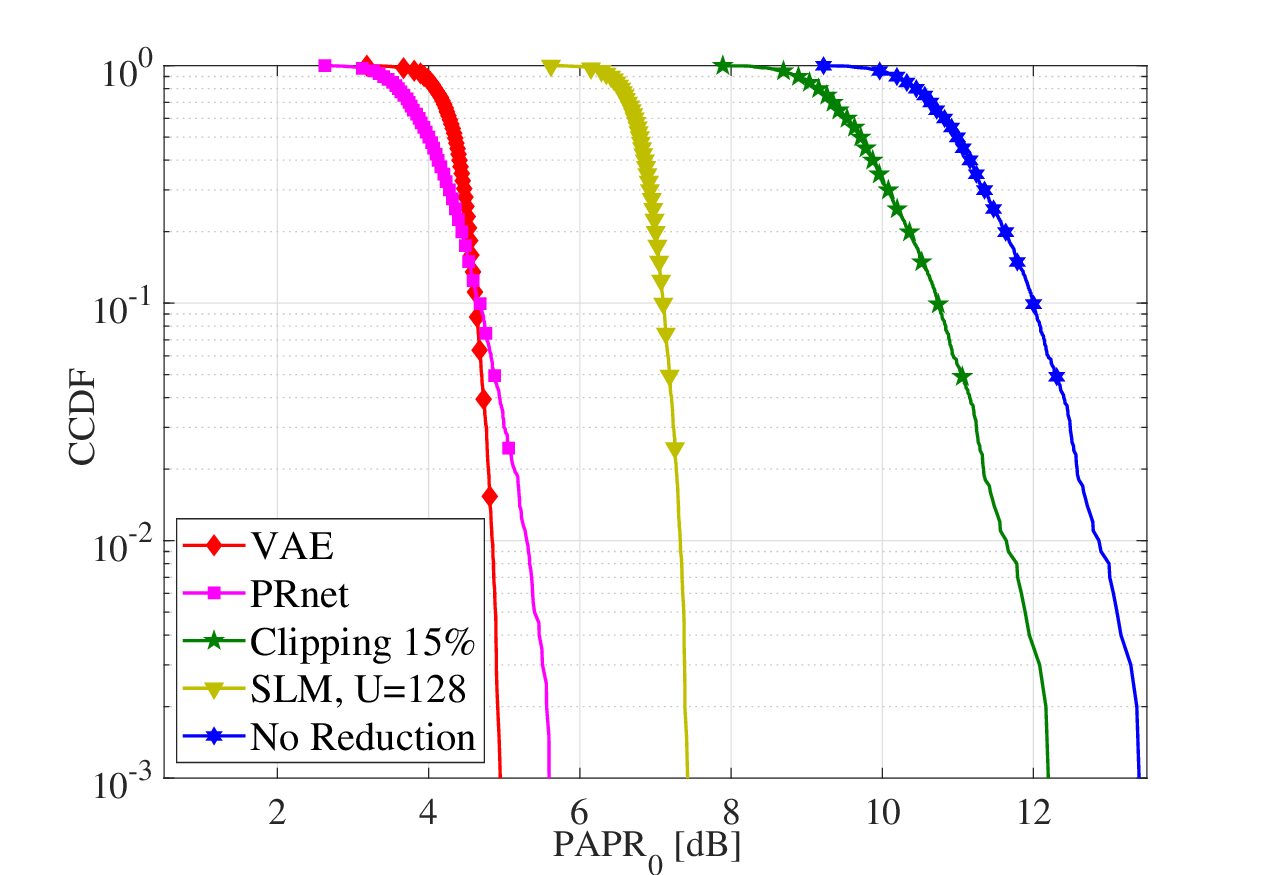}
      \caption{CCDF of PAPR.}
      \label{fig:PAPR-ACO}
     \end{subfigure}
     \caption{Comparison between the proposed method and other
techniques.}
\end{figure}

For the case of the 16-QAM constellation, Fig. \ref{fig:SER-ACO} shows the SER performance of the proposed method in comparison with other PAPR reduction techniques, including PRnet \cite{propose}, clipping, and SLM. It is seen that the VAE-based approach consistently achieves the lowest SER across all SNR values. This improvement of the proposed method stems from its ability to shape the constellation, which favors transmitting lower-energy symbols more frequently, thus enhancing communication robustness against noise.

Finally, the PAPR statistics of the proposed method and other techniques are illustrated in Fig.~\ref{fig:PAPR-ACO}, where the PAPR of each approach is calculated over 2000 random samples. It is observed that compared to non DL-based approaches (i.e., clipping and SLM), the proposed method exhibits a significant PAPR reduction. For instance, at the CCDF of $10^{-3}$, the proposed method reduces the PAPR by 2.5 dB and 7.1 dB compared to the SLM and clipping, respectively. Compared to the PRnet, although the proposed method performs worse at the low PAPR threshold (i.e., $\text{PAPR}_0 \leq 4.75$ dB), it outperforms at the higher threshold value. Specifically, at the CCDF of $10^{-3}$, the proposed method achieves a noticeable 0.5 dB PAPR reduction. 

\textit{Remarks on Computational Complexity}:
Let $N_I$, $N_O$, $N_L$, and $N_K$ be the number of input nodes, number of output nodes, number of layers, and number of nodes of hidden layers, respectively. Then, the number of multiplications of the proposed method is $N_IN_K+(N_L-2)N_K^2+N_KN_O$. The theoretical complexity of the proposed method is thus $\mathcal{O}$($N_IN_K+(N_L-2)N_K^2+N_KN_O$), where $\mathcal{O}(\cdot)$ denotes the Big O notation. The PRnet method has a complexity of $\mathcal{O}$($N_IN_K+(N_L-1)N_K^2$). The complexities of the clipping and SLM methods are $\mathcal{O}$($N_{\text{IFFT}}$) and $\mathcal{O}$($N_U N_{\text{IFFT}}\log N_{\text{IFFT}}$), respectively, where $N_{\text{IFFT}}$ and $N_U$ are the number of IFFT points and the number of phase sequences. We also evaluated the average computational times of the considered methods, which were implemented using Python on a Windows 11 desktop computer with an Intel Xeon E5 v3 processor. With the parameters given in TABLE I of the manuscript, the proposed method and the PRnet have average computational times of approximately 210 $\mu\text{s}$ and 180 $\mu\text{s}$, respectively. The computational time of the SLM method with $N_U=128$ is around 1270 $\mu\text{s}$, while the clipping method, due to its simplicity, takes only 10 $\mu\text{s}$.

Since OWC channels are either modeled as semi-static channels with channel coherence time being on the order of a few milliseconds or longer, the proposed method can obtain the optimal constellation shaping solution well within the channel coherence time.
\vspace{-0.3cm}
\section{Conclusions}
In this study, a VAE-based model for generating high-performance optical OFDM signals was presented. Simulation results demonstrated that the proposed signal design, by optimizing the probability and geometry of the constellation's symbols, achieved lower PAPR, lower SER, and higher MI compared to existing methods that employed uniform signaling. It is noted that while efficient OFDM waveform design has been extensively studied for the single-user configuration, little effort has been devoted to multi-user systems. In this regard, an AE-based waveform design for multi-user optical OFDM systems could be a promising research topic. 

%we employed a VAE model to enhance the performance of an ACO-OFDM system for optical signals, evaluated through three key metrics: SER, MI and PAPR. The simulation results demonstrate that the VAE-based approach outperforms conventional methods, highlighting its effectiveness in improving system reliability and power efficiency.
\bibliographystyle{ieeetr}
\bibliography{references}

\end{document}